\documentclass[11pt]{article}

\usepackage{color,graphicx}
\usepackage{young}
\usepackage[vcentermath]{youngtab}
\usepackage{amsmath,amssymb,graphicx}
\usepackage{hyperref}
\definecolor{darkred}{rgb}{0.65,0.15,0}
\hypersetup{pdfborder={0 0 0},colorlinks=true,urlcolor=darkred,citecolor=blue,linkcolor=darkred,linktocpage=true}

\usepackage{cite}
\usepackage{amsmath}
\usepackage{amsfonts}
\usepackage{amssymb}
\usepackage{graphicx}%
\usepackage{amsthm}
\usepackage{mathrsfs}
\usepackage[T1]{fontenc}
\usepackage{enumerate}
\usepackage{color}

\def\4diml{four-dimensional}

\def\-1{^{-1}}

\newcommand{\M}{\mathscr{M}}

\newcommand{\G}{\mathscr{G}}

\makeatletter

\@addtoreset{equation}{section}
\makeatother

\begin{document}

\thispagestyle{empty}

\vspace{5mm}

\begin{center}
{\LARGE \bf Generalized Supergravity Equations for the \\[2mm]WZW  Model}

\vspace{15mm}
\normalsize
{\large  Ali Eghbali\footnote{Corresponding author: eghbali978@gmail.com}, Simin Ghasemi-Sorkhabi\footnote{s.ghassemi.s@gmail.com}, Adel Rezaei-Aghdam\footnote{rezaei-a@azaruniv.ac.ir}
}

\vspace{2mm}
{\small \em Department of Physics, Faculty of Basic Sciences,\\
Azarbaijan Shahid Madani University, 53714-161, Tabriz, Iran}\\

\vspace{7mm}


\vspace{6mm}

\begin{tabular}{p{12cm}}
{\small
Generalized supergravity equations (GSEs) were originally proposed as a modification of the standard IIB supergravity equations to
satisfy the background of $\eta$-deformed $AdS_5 \times S^5$ introduced by Arutyunov {\it et al}. In this study, we proceed to write down the
GSEs for the Wess-Zumino-Witten (WZW) models based on Lie groups.
First, we simplify the GSEs for the WZW model by imposing the conditions for vanishing of the one-loop beta function equations.
Then, by introducing an {\it Ansatz} for the Killing vector field $I$, it is shown that the Killing equation ${\cal L}_{_{I}} G_{_{\mu \nu}}=0$ is held.
In addition, we introduce a generalized Killing vector such that the existence of a solution to the GSEs requires that this vector be light-like.
In this way, we solve the simplified GSEs for the WZW models constructed on Lie groups up to dimension five.
Unfortunately, two of the groups do not have light-like vectors and therefore do not admit the GSEs.
}
\end{tabular}
\vspace{-1mm}
\end{center}


\setcounter{page}{1}
\newpage
\tableofcontents


\section{\label{Sec.I} Introduction}

About a decade has passed since the introduction of the GSEs of a prescription invented by Arutyunov {\it et al}
\cite{Arutyunov1}.
The most prominent feature of which is the absence of a scalar dilaton that this is one of the main
differences between the GSEs and usual supergravity equations,
although on-shell supergravity configurations are special solutions to the GSEs.
After that, Tseytlin and Wulff demonstrated that the $\kappa$-symmetry
constraints of the generalized supergravity (GS) formulation of type IIB superstring on an arbitrary background leads to
the GSEs including an extra Killing vector field $I$, rather than the usual supergravities \cite{Wulff3}.
Then, with the aim of constructing a suitable counterterm  (generalized Fradkin-Tseytlin counterterm)
for background fields satisfying the GSEs, the calculations of Tseytlin and Wulff were generalized \cite{w.muck}.
After publication of \cite{Wulff3} it became clear that integrable deformations should satisfy
the GSEs instead of usual supergravity equations.
Since the introduction of the Yang-Baxter deformation \cite{Klimcik11} as a systematic way of generating such integrable deformations, this topic
led to the intriguing discovery the GSEs, a new low-energy effective theory (see \cite{Orlando.Yoshida} for an interesting review).
In fact, the GSEs as the generalization of the type IIB supergravity were found
in the context of integrable deformations of the $AdS_5 \times S^5$  type II superstring sigma model \cite{{Delduc1},{Delduc2},{Yoshida1},{Wulff1}}
which are closely related to non-Abelian T-duality transformations \cite{{Yoshida2},{Hoare1},{Wulff2}}.
Indeed, one can find the physical interpretation of the vector field $I$ by employing the Yang-Baxter deformation,
because this can be regarded as a solution generation technique in GS \cite{sheikh1} (see, also, \cite{{sheikh2},{sheikh3}}).
It should be noted that the classical action corresponding to the GSEs has not been revealed yet, and
only the equations of motion are presented.
In Ref. \cite{Arutyunov1}, it has been shown that there is a T-duality transformation rule from a solution of the GSEs to a
solution of usual supergravity. On the other hand, one can map a solution of
usual supergravity with a linear dilaton to a solution of the GSEs by performing a
formal T-duality transformation along a direction for which the dilaton is not isometric \cite{Hoare2}.
These results indicate that solutions of usual supergravity and the GSEs should be treated
on an equal footing in the context of string theory, because the T-duality is a symmetry of
string theory.

Later, the GSEs attracted the attention of many researchers interested in this field.
In \cite{yuho1}, it was shown that the bosonic part of the GSEs can be completely reproduced from the modified double field theory; moreover,
when the dilaton has a linear dual-coordinate dependence, the equations of motion of the double field theory lead to the
GSEs. However, we think that the results presented in \cite{yuho1} have provided positive evidence
that superstring theories defined on solutions of the GSEs are
Weyl invariant (see, also, \cite{yuho2}).  Recently, a local counterterm has constructed in \cite{yuho3} that cancels
out the Weyl anomaly, and string theory may be consistently defined even in the generalized background.
In \cite{M.Y.E.Colgain}, it has obtained the non-Abelian T-dual spaces corresponding to some of the Bianchi cosmologies and then have shown that dual backgrounds indeed satisfy the GSEs.
Additionally, in \cite{hlavaty1} by applying Poisson-Lie T-plurality \cite{vonUnge} on Bianchi cosmologies, it has presented a formula for the vector field $I$
as well as transformation rule for dilaton field. Then, it has shown there that plural backgrounds together with the introduced dilaton field and $I$ satisfy the GSEs (see, also, \cite{{hlavaty2},{hlavaty3}}).
Recently, we have investigated the solutions of the GSEs in three dimensions with the metric of BTZ black hole, in such a way that
only the cases with $J = M = 0$ and $J = 0, M \neq 0$ of the BTZ metric satisfied the GSEs \cite{TTMP.eghr}.
There, we have also shown that the charged black string solution found by Horne
and Horowitz \cite{Horowitz1}, which is Abelian T-dual to the BTZ black
hole solution, can be considered as a solution for the GSEs.

The main goal of this paper is to write down the GSEs for the WZW models based on Lie groups.
After the Introduction section,  we give a short review of the GSEs, where necessary formulas are summarized.
We impose the conditions for vanishing of the one-loop beta function equations
to simplify the GSEs for the WZW model. Then, we introduce an {\it Ansatz} for the vector field $I$, in such a way that it is proved that
the Killing equation ${\cal L}_{_{I}} G_{_{\mu \nu}}=0$ is held.
Most importantly, a generalized Killing vector is introduced such that the necessary condition for having a solution for the GSEs
is that this vector is light-like.
Finally, we solve the simplified GSEs for the WZW models constructed on real Lie groups up to dimension five.
Among these models only models built on the $SO(3)$ and $SO(3) \otimes SO(2)$ Lie groups do not admit the GSEs.
As will be shown, the reason behind this is that these groups do not have light-like vectors.
We conclude with a final discussion of the results with remarks and perspectives.

\section{\label{Sec.II} Conditions on which WZW model can be considered as a solution for the GSEs}

Before proceeding to find the condition on which WZW model can be considered as a solution of the GSEs, let us give
a short review of the GSEs.

\subsection{ A short review of the GSEs}

In the absence of the Ramond-Ramond fields, the set of the bosonic GSEs in D dimensions take the following form
(here fermionic fields are set to zero) \cite{Arutyunov1}:
\begin{eqnarray}
{\cal R}_{{\mu \nu}}-\frac{1}{4} H_{\mu \rho\sigma} {H^{\rho\sigma}}_{\nu}+{\nabla}_\mu X_\nu +{\nabla}_\nu X_\mu  &=&0,\label{2.1}\\
\frac{1}{2}{\nabla}^\lambda H_{{\lambda \mu \nu}} -X^{\lambda} H_{{\lambda \mu \nu}}-{\nabla}_\mu X_\nu +{\nabla}_\nu X_\mu &=&0,\label{2.2}\\
{\cal R}-\frac{1}{12}  H^{{2}} +4 {\nabla}_\mu X^\mu -4 X_\mu X^\mu-4 \Lambda&=&0,\label{2.3}
\end{eqnarray}
where ${\cal R}_{{\mu \nu}}$ and ${\cal R}$ are the respective Ricci tensor and scalar curvature that are calculated from the metric ${G}_{{\mu \nu}}$,
and $\Lambda$ is the cosmological constant. Here, the $D$-dimensional indices $\mu, \nu,...$ of coordinates
$x^\mu$ of manifold $\M$ are raised or lowered with the metric ${G}_{{\mu \nu}}$. The covariant derivative
${\nabla}_\mu$ is the conventional Levi-Civita connection associated with   ${G}_{{\mu \nu}}$.
A vector field $I=I^{^\mu} \partial_\mu$  and a one-form $Z=Z_\mu dx^\mu$  are defined so as to satisfy
\begin{eqnarray}
{\cal L}_{_{I}} G_{_{\mu \nu}}&=&0,\label{2.4}\\
{\cal L}_{_{I}} B_{_{\mu \nu}}&=&0,\label{2.5}\\
{\nabla}_\mu Z_\nu -{\nabla}_\nu Z_\mu +I^\lambda H_{{\lambda \mu \nu}}  &=&0,\label{2.6}\\
I^\mu Z_\mu&=&0,\label{2.7}
\end{eqnarray}
where ${\cal L}$ stands for the Lie derivative. Moreover, the $X_{_\mu}$ are related to the one-form $Z$ via $X_\mu= I_\mu + Z_\mu$.
 The field strength $H_{_{\mu \nu \rho}}$  corresponding to anti-symmetric tensor field $B_{_{\mu \nu}}$ ($B$-field) is defined as
\begin{eqnarray}\label{2.8}
H_{_{\mu \nu \rho}}= \partial_\mu B_{_{\nu \rho}}+\partial_\nu B_{_{\rho \mu}}+\partial_\rho B_{_{\mu \nu}}.
\end{eqnarray}
Furthermore, the conventional dilaton is included in $Z_\mu$  as follows:
\begin{eqnarray}\label{2.9}
Z_\mu= \partial_\mu \Phi + B_{_{\nu \mu}} I^{^{\nu}},
\end{eqnarray}
where  $\Phi$ is a scalar dilaton field.
Notice that one may use \eqref{2.9} to rewrite  \eqref{2.7} in the form of $I^\mu~ \partial_\mu \Phi =0$.
A remarkable point is that if we set $I^{^{\mu}} = 0$, then one gets that $X_\mu= \partial_\mu \Phi$, and thus, the
GSEs reduce to the usual supergravity equations (corresponding to the one-loop beta function equations in sigma model language \cite{callan}).

\subsection{ The GSEs for the WZW model}
The main purpose of this subsection is to find under which conditions, the background of WZW model can satisfy the GSEs.
This will allow us to introduce new standpoint of the WZW model in the GSEs.
Before proceeding to doing, let us review the WZW model setting based on a Lie group $G$.

\subsubsection{\it The WZW model}
The WZW model based on a Lie group $G$ can be expressed as \cite{witten1}
\begin{eqnarray}\label{2.10}
S_{_{WZW}}(g) &=&  \frac{1}{2} \int_{_\Sigma} d\sigma^+ d\sigma^-\;{\Omega}_{ij}
			L^{i}_{+}\;
			L^{j}_{-}
			+\frac{1}{12} \int_{_B} d^3 \sigma~
			\varepsilon^{ \gamma \alpha \beta}~{\Omega}_{ik} \;f_{jl}^{~~k}~
			L^{i}_{_\gamma} L^{j}_{_\alpha} L^{l}_{_\beta},
\end{eqnarray}
where $\sigma^{\alpha}=(\sigma^+ , \sigma^-)$ are the standard light-cone variables on the worldsheet, which are defined in terms of the worldsheet coordinates by
$\sigma^{\pm} =(\tau \pm \sigma)/2$ together with $\partial_{_\pm}=\partial_{\tau} \pm \partial_{\sigma}$, and $\varepsilon^{ \gamma \alpha \beta}$ is the Levi-Civita symbol.
The integrations are over worldsheet $\Sigma$ and a 3-dimensional manifold with boundary $\partial B = \Sigma$, respectively,
and $L^{i}_{\alpha}$'s are the components of the left-invariant one-forms on $G$
which are constructed by means of an element $g: \Sigma \rightarrow G$ in the following way
\begin{eqnarray}\label{2.11}
g^{-1} \partial_{_\alpha} g = L^{i}_{\alpha} T_i=\partial_{_\alpha} x^\mu~L^{~i}_{\mu}~T_i,
\end{eqnarray}
where $x^\mu$'s represent the coordinates of the group manifold, and $T_i$'s, $i=1, \cdots, dim~G$ are the basis of the Lie algebra ${\cal G}$ of $G$.
The inverse of $L^{~~i}_{\mu}$ will be denoted by $L^{\mu}_{~~i}$ and for notational convenience we will also denote the transpose of the inverse of $L^{~~i}_{\mu}$ by $(L^{\mu}_{~~i})^t$.
We note that $\Omega_{ij}$ defined by $\Omega_{ij} =<T_i , T_j>$ is a non-degenerate ad-invariant symmetric bilinear form on the Lie algebra ${\G}$
with structure constants $f_{ij}^{~~k}$ which satisfies the following relation \cite{witten}
\begin{eqnarray}\label{2.13}
f_{ij}^{\;\;k} \;\Omega_{kl}+ f_{il}^{\;\;k} \;\Omega_{kj}\;=\;0.
\end{eqnarray}
By regarding the WZW action \eqref{2.10} as a 2-dimensional non-linear sigma model of the form $S=1/2~\int \!d\sigma^+ d\sigma^- (G_{_{\mu\nu}}
+B_{_{\mu\nu}})\partial_{_+}x{^\mu} \partial_{_-}x^{\nu}$, one can derive
the corresponding line element and $B$-field, giving us
\begin{eqnarray}\label{2.14}
ds^2 &=& G_{_{\mu\nu}} dx{^\mu}   dx^{\nu} = L_\mu^{~i}~L_\nu^{~j}~\Omega_{ij} dx{^\mu}   dx^{\nu},\nonumber\\
B &=& \frac{1}{2} B_{_{\mu\nu}} dx{^\mu}  \wedge dx^{\nu},
\end{eqnarray}
such that the field strength of the field  $B_{_{\mu\nu}}$ is given by
\begin{eqnarray}
H_{_{\mu \nu \rho}}= \Omega_{ik} f_{jl}^{~~k}  ~L_\mu^{~i}~L_\nu^{~j}~L_\rho^{~l}.\label{2.14.1}
\end{eqnarray}

\subsubsection{\it Simplifying the GSEs for the WZW model}
Let us now turn our attention to the GSEs for the WZW model. To begin, one may use the definition $X_\mu= I_\mu + Z_\mu$ together with formula \eqref{2.9}
to rewrite down equations \eqref{2.1}-\eqref{2.7} in the following form
\begin{eqnarray}
\Big[{\cal R}_{{\mu \nu}}-\frac{1}{4} H_{\mu \rho\sigma} {H^{\rho\sigma}}_{\nu}+2{\nabla}_\mu {\nabla}_\nu \Phi\Big] +({\nabla}_\nu \omega_\mu+{\nabla}_\mu \omega_\nu)  &=&0,\label{2.15}\\
\Big[\frac{1}{2}{\nabla^\lambda} H_{{\lambda \mu \nu}}-{\nabla^\lambda}\Phi~H_{{\lambda \mu \nu}}\Big]+(-\omega^{\lambda} H_{{\lambda \mu \nu}}+{\nabla}_\nu \omega_\mu -{\nabla}_\mu \omega_\nu) &=&0,\label{2.16}\\
\Big[-4 \Lambda +\frac{1}{6}  H^{{2}} +2 {\nabla}^2 \Phi-4 (\nabla \Phi)^2 \Big]+(2 \nabla_\mu \omega ^\mu -8 \nabla_\mu \Phi ~\omega ^\mu -4 ~\omega_\mu ~ \omega ^\mu) &=&0,\label{2.17}\\
\partial_{\mu} I^{\lambda} ~ G_{_{\lambda \nu}} +I^{\lambda} \partial_{\lambda} G_{_{\mu \nu}}+\partial_{\nu} I^{\lambda} ~ G_{_{\mu \lambda}}&=&0,\label{2.18}\\
\partial_{\mu} I^{\lambda} ~ B_{_{\lambda \nu}} +I^{\lambda} \partial_{\lambda} B_{_{\mu \nu}}+\partial_{\nu} I^{\lambda} ~ B_{_{\mu \lambda}}&=&0,\label{2.19}\\
{\nabla}_\mu (B_{\sigma \nu}~I^\sigma)-{\nabla}_\nu (B_{\sigma \mu}~I^\sigma)+I^\lambda H_{{\lambda \mu \nu}}  &=&0,\label{2.20}\\
I^\mu~ \partial_\mu \Phi &=&0,\label{2.21}
\end{eqnarray}
where we have defined
\begin{eqnarray}
\omega_\mu = I_\mu +B_{\nu \mu}I^\nu=(G_{\nu \mu} +B_{\nu \mu})I^\nu.\label{2.22}
\end{eqnarray}
Obviously, the terms in brackets of equations \eqref{2.15}-\eqref{2.17} are nothing but the one-loop beta function equations. On the other hand,
as known, the WZW model should be conformally invariant, namely, one can look at the vanishing of the one-loop beta function equations and
see that the background of WZW model always satisfies these equation with a constant dilaton field.
Accordingly, equations \eqref{2.15}-\eqref{2.20} can be, respectively, expressed as\footnote{One may use \eqref{2.26} instead of \eqref{2.18}.}
\begin{eqnarray}
{\nabla}_\mu \omega_\nu +{\nabla}_\nu \omega_\mu  &=&0,\label{2.23}\\
-\omega^{\lambda} H_{{\lambda \mu \nu}}+{\nabla}_\nu \omega_\mu -{\nabla}_\mu \omega_\nu &=&0,\label{2.24}\\
\nabla_\mu \omega^\mu  -2 ~\omega_\mu ~ \omega ^\mu&=&0,\label{2.25}\\
{\nabla}_\mu I_{\nu} +{\nabla}_\nu I_{\mu} &=&0,\label{2.26}\\
\partial_{\mu} I^{\lambda} ~ B_{_{\lambda \nu}} +I^{\lambda} \partial_{\lambda} B_{_{\mu \nu}}+\partial_{\nu} I^{\lambda} ~ B_{_{\mu \lambda}}&=&0,\label{2.27}\\
{\nabla}_\mu (B_{_{\rho \nu}} I^\rho) -{\nabla}_\nu  (B_{_{\rho \mu}} I^\rho) +I^\lambda H_{{\lambda \mu \nu}}  &=&0.\label{2.28}
\end{eqnarray}
Note that equation \eqref{2.21} holds with a constant dilton field. Indeed, equation \eqref{2.23} states that $ \omega^\mu =G^{\mu \nu}  \omega_\nu$ is a Killing vector, which we call a {\it generalized Killing vector}.

In the following, we would like to simplify the above equations and then write them in terms of three fields $(G_{\mu \nu} , B_{\mu \nu}, I^{^\mu})$.
First, by multiplying the sides of \eqref{2.23} in $G^{\mu \nu}$ we obtain
\begin{eqnarray}\label{2.30}
\nabla_\mu \omega ^\mu =0,
\end{eqnarray}
by substituting \eqref{2.30} into \eqref{2.25} one gets
\begin{eqnarray}\label{2.31}
\omega_\mu ~\omega ^\mu =0.
\end{eqnarray}
This condition states that $\omega^\mu$ is a light-like vector.
Then, by using the definition \eqref{2.22} one may write \eqref{2.31} in the following form
\begin{eqnarray}\label{2.32}
I_\mu ~I^\mu+B_{\mu \nu}~B^{\mu \gamma}~I^\nu~I_\gamma=0.
\end{eqnarray}
This equation can be an appropriate substitute for \eqref{2.25}. We also employ the definition of $\omega_{\mu}$ of equation \eqref{2.22} together with
\eqref{2.26} to write equation \eqref{2.23} as follows:
\begin{eqnarray}\label{2.33}
\nabla_\mu (B_{\gamma \nu}~I^\gamma)+\nabla_\nu (B_{\gamma \mu}~I^\gamma)=0.
\end{eqnarray}
Equation \eqref{2.33} helps us to write equation \eqref{2.28} as
\begin{eqnarray}\label{2.34}
\nabla_\mu (B_{\gamma \nu}~I^{\gamma})=-\frac{1}{2}~I^{\lambda}~H_{\lambda \mu \nu},
\end{eqnarray}
by employing \eqref{2.8} and then by utilizing \eqref{2.27} in the form of
$I^\lambda ~\partial_{\lambda}B_{\mu \nu}= -\partial_{\mu}I^\lambda~B_{\lambda \nu}-\partial_{\nu}I^\lambda~B_{\mu \lambda }$,
one concludes that equations \eqref{2.34} and \eqref{2.33} are identical.
 Let us now look at equation \eqref{2.24}. By substituting the definition \eqref{2.22} into \eqref{2.24} and then by employing
equation \eqref{2.34} we find that
\begin{eqnarray}\label{2.35}
{\nabla}_\mu I_\nu - {\nabla}_\nu I_\mu - G^{\lambda \sigma} ~B_{ \sigma \gamma}~I^\gamma~H_{\lambda \mu  \nu} =0.
\end{eqnarray}
From contraction the above equation with \eqref{2.26} one arrives at
\begin{eqnarray}\label{2.36}
{\nabla}_\mu I_\nu =\frac{1}{2}  G^{\lambda \sigma} ~B_{ \sigma \gamma}~I^\gamma~H_{\lambda \mu  \nu}.
\end{eqnarray}
Finally, after the simplifications made above, we conclude that only the following equations are valuable
\begin{eqnarray}
{\nabla}_\mu I_\nu -\frac{1}{2}  G^{\lambda \sigma} ~B_{ \sigma \gamma}~I^\gamma~H_{\lambda \mu  \nu} &=& 0, \label{2.37}\\
I_\mu~I^\mu+B_{\mu \nu}~B^{\mu \gamma}~I^\nu~I_{\gamma} &=&0,\label{2.38}\\
\nabla _\mu I_\nu+\nabla _\nu I_\mu &=&0,\label{2.39}\\
\partial_{\mu} I^{\lambda} ~ B_{_{\lambda \nu}} +I^{\lambda} \partial_{\lambda} B_{_{\mu \nu}}+\partial_{\nu} I^{\lambda} ~ B_{_{\mu \lambda}}&=&0,\label{2.40}\\
\nabla_\mu (B_{\gamma \nu}~I^{\gamma})+\frac{1}{2}~I^{\lambda}~H_{\lambda \mu \nu} &=&0.\label{2.41}
\end{eqnarray}
Now, a question which arises is whether the above equations hold for any WZW model constructed out on a Lie group $G$. We note that in examining the above equations for
a background of the WZW model including two fields $(G_{\mu \nu} , B_{\mu \nu})$,
only the vector field $I$ is unknown.
In order to answer the question raised above, we start by introducing the following {\it Ansatz} for the vector field $I$
\begin{eqnarray}
I^\mu=(L^{\mu}_{~~i})^t~I^i, \label{2.42}
\end{eqnarray}
for some constant values of $I^i$.
As we will show below, this {\it Ansatz} solves equation \eqref{2.39}.
First, by inserting \eqref{2.42} in equation \eqref{2.39} or equivalently in \eqref{2.18},
and then by employing the first equation of \eqref{2.14} we get
\begin{eqnarray}
&&\Big[\partial_{\mu} (L^{\lambda}_{~~i})^t ~ L^{~~j}_{\lambda} L^{~~k}_{\nu}
+ \partial_{\nu} (L^{\lambda}_{~~i})^t ~ L^{~~j}_{\mu} L^{~~k}_{\lambda}~~~~~~~~~~~~\nonumber\\
&&~~~~~~~~~~~~~~~~~~~+ \partial_{\lambda}L^{~~j}_{\mu}~  (L^{\lambda}_{~~i})^t L^{~~k}_{\nu}
+ \partial_{\lambda}L^{~~k}_{\mu}~  (L^{\lambda}_{~~i})^t L^{~~j}_{\mu}\Big]\Omega_{jk} I^i =0, \label{2.43}
\end{eqnarray}
by multiplying the sides of the resulting equation in $L^{\mu}_{~~l} (L^{\nu}_{~~m})^t$ one obtains that
\begin{eqnarray}
&&\big[\partial_{\mu} (L^{\lambda}_{~~i})^t ~ L^{~~j}_{\lambda}+ \partial_{\lambda}L^{~~j}_{\mu}~(L^{\lambda}_{~~i})^t\big] L^{\mu}_{~~l}~ \Omega_{jm} I^i\nonumber\\
&&~~~~~~~~~~~~~~~~~~~~~+\big[\partial_{\nu} (L^{\lambda}_{~~i})^t ~  L^{~~k}_{\lambda}+
\partial_{\lambda}L^{~~k}_{\nu}~  (L^{\lambda}_{~~i})^t \big] (L^{\nu}_{~~m})^t~ \Omega_{lk} I^i =0. \label{2.44}
\end{eqnarray}
After some algebraic manipulation and finally by using the Maurer-Cartan equation
$\partial_{\mu} L^{~~i}_{\nu} - \partial_{\nu} L^{~~i}_{\mu}=-f^i_{~jk}  L^{~~j}_{\mu}  L^{~~k}_{\nu} $
we arrive at the ad-invariance condition on the algebraic metric, \eqref{2.13}.
Therefore, by considering condition \eqref{2.42}, equation \eqref{2.39} can be removed from equations \eqref{2.37}-\eqref{2.41}.
In the next section, we use \eqref{2.42} to solve the rest of the equations for the WZW models constructed on real Lie groups up to dimension five.


\section{\label{Sec.III} Examples of WZW models on real Lie groups up to dimension five as solutions of the GSEs}

All real Lie algebras with dimension up to five are known \cite{patera}. In \cite{kehagias1}, it has been examined for which of these an
ad-invariant, symmetric and non-degenerate metric exists.
The first non-trivial case is the well-known three-dimensional $sl(2 , \mathbb{R})$ and $so(3)$, and
in four dimensions the centrally extended Euclidean algebra of the group $E_2^c$
that is isomorphic to the $A_{4,10}$ Lie algebra of Patera's classification \cite{patera},
the Heisenberg Lie algebra $h_4$ (isomorphic to the $A_{4,8}$), $gl(2 , \mathbb{R})=sl(2 , \mathbb{R}) \oplus so(2)$
and $so(3) \oplus so(2)$.
In five dimensions there exist only one such algebra, denoted by $A_{5,3}$ for which corresponding WZW model was constructed in \cite{kehagias1}.
Nappi and Witten \cite{witten} constructed the action of WZW for the non-semisimple group $E_2^c$ such that
their model described string propagation on a four-dimensional
space-time in the background of a gravitational plane wave.
Subsequently, by expanding this structure to other non-semisimple Lie groups \cite{Sfetsos1},
the WZW model on the Heisenberg group with arbitrary dimension (especially $H_{4}$) was introduced by Kehagias and Meessen \cite{kehagias2}
(see, also, \cite{{eghbali11},{Eghbali.2}}).
For the WZW models based on the three-dimensional Lie groups one can refer to \cite{{Alvarez-Gaume},{Alekseev1},{Maldacena},{EMR13}}.

As was mentioned earlier, by using the {\it Ansatz} \eqref{2.42} we shall solve the simplified equations
of GSEs (equations \eqref{2.37}, \eqref{2.38}, \eqref{2.40} and \eqref{2.41}) for the WZW models constructed on Lie groups up to dimension five.
The case of $H_4$ is precisely interpreted.
It is also interesting to see that the cases $SO(3)$ and $SO(3) \otimes SO(2)$ out of all the cases mentioned above does satisfy the GSEs with $I^\mu =0$.
Now, some questions arise. What is unique about these model? Could it be explained in terms of the properties of their Killing form?
Is it because there are no light-like vectors in these models?
In order to answer these questions, we focus on equation \eqref{2.31}, which is a necessary condition for the existence of a solution of the GSEs.
Calculating the vectors $I$ from equation \eqref{2.42},
it can be easily shown that there is no light-like generalized Killing vector
for the $SO(3)$ and $SO(3) \otimes SO(2)$, but rather in both cases we have $\omega_{{\mu}} \omega^{{\mu}} >0$. Accordingly, the models on these groups do not admit the GSEs.

\subsection{\it The case of $H_{4}$}

Before proceeding to write down the WZW  model based on the $H_{4}$ Lie group, let us introduce the $h_4$ Lie algebra of the $H_4$.
It is defined by the set of generators $(T_{1}, T_{2}, T_{3}, T_{4})$ with the following non-zero Lie brackets
\begin{eqnarray}\label{3.1}
[T_{1} , T_{2}]~=~T_{2},~~~~~[T_{1} ,T_{3}]~=~-T_{3},~~~~~[T_{3} , T_{2}]~=~T_{4}.
\end{eqnarray}
The $h_4$ Lie algebra possesses a non-degenerate metric which can be obtained by inserting
the structure constants of \eqref{3.1} into equation \eqref{2.13}, giving \cite{eghbali11}
\begin{eqnarray}\label{3.2}
\Omega_{ij}=\left( \begin{tabular}{cccc}
                 $\rho$ & 0 & 0 & -$1$ \\
                 0 & 0 & $1$ & 0 \\
                 0 & $1$& 0 & 0 \\
                 -$1$ & 0 & 0 & 0 \\
                 \end{tabular} \right),
\end{eqnarray}
where $\rho$ is an arbitrary constant. In order to calculate the left-invariant one-forms, we parameterize an element of the $H_4$ as
\begin{eqnarray}\label{3.3}
g=e^{v T_{_4}} ~ e^{u T_{_3}} ~e^{x T_{_1}} ~e^{y T_{_2}},
\end{eqnarray}
where $x^{\mu} =(x, y, u, v)$ stand for the coordinates of the group manifold.
Inserting \eqref{3.3} into \eqref{2.11} one can obtain the matrix form of the corresponding left-invariant one-forms components, giving
\begin{eqnarray}\label{3.4}
L^{~~i}_{\mu}=\left( \begin{tabular}{cccc}
                 1 & $y$ & 0 & 0 \\
                 0 & 1 & 0 & 0 \\
                 0 & 0 & $e^x$ & $y e^x$ \\
                 0 & 0 & 0 & 1 \\
                 \end{tabular} \right).
\end{eqnarray}
Putting these pieces together, one can find the WZW action on the $H_4$. Then, by
identifying the resulting WZW action with the standard form of the sigma model (introduced in section \ref{Sec.II})
and finally by using \eqref{2.14}
one can read off the line element and $B$-field, giving us \cite{eghbali11}
\begin{eqnarray}
ds^2 &=&\rho~d x^{2}-2~ dx~ dv + 2 e^x~ dy~ du,\label{3.5}\\
B&=&-ye^x~ dx \wedge du.\label{3.6}
\end{eqnarray}
Now, we employ formula \eqref{2.42} to write the vector field $I$ for the present model. Using \eqref{3.4} and then
considering $I^i$ in the form of
$I^i=\left( \begin{tabular}{c}
                 $\beta_{_1}$ \\
                 $\beta_{_2}$ \\
                $\beta_{_3}$ \\
                $\beta_{_4}$ \\
                 \end{tabular} \right)$, for some constants $\beta_{_i}$, we get
\begin{eqnarray}\label{3.7}
I= I^\mu \partial_{_\mu} = \beta_{_1} ~\partial_{_x}+(-y\beta_{_1}+\beta_{_2})\partial_{_y}+ \beta_{_3}~e^{-x} \partial_{_u}+
(-\beta_{_3} y+\beta_{_4})\partial_{_v}.
\end{eqnarray}


\begin{center}
		\small {{{\bf Table 1.}~ The WZW models on real Lie groups up to dimension five as solutions of the GSEs}}
		{\scriptsize
			\renewcommand{\arraystretch}{1.35}{
			\begin{tabular}{p{1.9cm}llll} \hline \hline
Lie group &  Left-invariant & Background including  &  Vector field $I$\\
\vspace{-3mm}
 &   one-forms & the metric and $B$-field & \\ \hline
{$SL(2,\mathbb{R})$} & $ L^{1}_{\pm} = u e^{-2\phi} \partial_{_\pm} y +  \partial_{_\pm} \phi,$ &  $ds^{2}=d\phi^2+e^{-2\phi}~du~dy,$ & \\
  &  $ L^{2}_{\pm} = e^{-2\phi} \partial_{_\pm} y,$ &  $B=-\frac{1}{2}~e^{-2\phi}~du\wedge dy$ & $I=\beta_{_3}~\partial_{_u}$\\
  &  $ L^{3}_{\pm} = -u^2 e^{-2\phi} \partial_{_\pm} y$ &   & \\
  &  $ ~~~~~~ -2u \partial_{_\pm} \phi + \partial_{_\pm} u,$ &   & \\
\\

{${SO(3)}^{\ast}$} & ${L^{1}_{\pm}} = \frac{1}{2} \sin\omega \cos\bar{\phi} (-\partial_{_\pm} \varphi_{_1} $  & $ds^{2}=\cos^2 {\frac{\omega}{2}}d\varphi_{_1}^2+\sin^2 {\frac{\omega}{2}}d\varphi_{_2}^2 + d{\omega}^2,$ & \\
  & $~~~~~+\partial_{_\pm} \varphi_{_2}) +\sin\bar{\phi} \partial_{_\pm} \omega,$ &  $B=-\frac{1}{2} \cos \omega~d\varphi_{_1} \wedge d\varphi_{_2}$ & $I=0$\\
  & $L^{2}_{\pm} = \frac{1}{2} \sin\omega \sin\bar{\phi} (\partial_{_\pm} \varphi_{_1} $ &     & \\
  & $~~~~~-\partial_{_\pm} \varphi_{_2}) +\cos\bar{\phi} \partial_{_\pm} \omega,$ &   & \\
  & $L^{3}_{\pm} =  \cos^2\frac{\omega}{2} \partial_{_\pm} \varphi_{_1} + \sin^2\frac{\omega}{2} \partial_{_\pm} \varphi_{_2}$ &     & \\\\

  {${SO(3)}\otimes SO(2)$} & ${L^{1}_{\pm}} = \frac{1}{2} \sin\omega \cos\bar{\phi} (-\partial_{_\pm} \varphi_{_1} $  & $ds^{2}=\cos^2 {\frac{\omega}{2}}d\varphi_{_1}^2+\sin^2 {\frac{\omega}{2}}d\varphi_{_2}^2
  + d{\omega}^2 +b  du^2,$ & \\
  & $~~~~~+\partial_{_\pm} \varphi_{_2}) +\sin\bar{\phi} \partial_{_\pm} \omega,$ &  $B=-\frac{1}{2} \cos \omega~d\varphi_{_1} \wedge d\varphi_{_2}$ & $I=0$\\
  & $L^{2}_{\pm} = \frac{1}{2} \sin\omega \sin\bar{\phi} (\partial_{_\pm} \varphi_{_1} $ &     & \\
  & $~~~~~-\partial_{_\pm} \varphi_{_2}) +\cos\bar{\phi} \partial_{_\pm} \omega,$ &   & \\
  & $L^{3}_{\pm} =  \cos^2\frac{\omega}{2} \partial_{_\pm} \varphi_{_1} + \sin^2\frac{\omega}{2} \partial_{_\pm} \varphi_{_2},$ &     & \\
  & $L^{4}_{\pm} =   \partial_{_\pm} u$ &     & \\

\\
{$H_4$} & $ L^{1}_{\pm} =  \partial_{_\pm} x,$ & $ds^{2}=\rho~dx^2-2~dx~dv+2e^x~dy~du,$ & \\
& $ L^{2}_{\pm} = y \partial_{_\pm} x +  \partial_{_\pm} y,$ & $B=y~e^x~du\wedge dx$ & $I=\beta_{_4}\partial_{_v}$\\
& $ L^{3}_{\pm} = e^{x} \partial_{_\pm} u,$ &  & \\
& $ L^{4}_{\pm} = y e^{x} \partial_{_\pm} u +  \partial_{_\pm} v$ &  & \\

\\
{$A_{4,10}$} & $ L^{1}_{\pm} = \cos u \partial_{_\pm} a_{_1} +\sin u \partial_{_\pm} a_{_2},$ & $ds^{2}=da_{_1}^2+da_{_2}^2+b~du^2+a_{_1}~da_{_2}~du$ & \\
  & $ L^{2}_{\pm} = \cos u \partial_{_\pm} a_{_2} -\sin u \partial_{_\pm} a_{_1},$&  ~~~~~~~$-a_{_2}~da_{_1}~du+2~du~dv,$ & \\
  & $ L^{3}_{\pm} = \partial_{_\pm} u,$ & $B=u~da_{_1}\wedge da_{_2}$ & $I=\beta_{_4} \partial_{_v}$\\
 & $ L^{4}_{\pm} = \frac{1}{2} (a_{_2}\partial_{_\pm} a_{_1} - a_{_1}\partial_{_\pm} a_{_2}) +\partial_{_\pm} v$ &  & \\
\\

 {$GL(2,\mathbb{R})$} & $ L^{1}_{\pm} = u e^{-2x} \partial_{_\pm} y +\partial_{_\pm} x,$ & $ds^{2}=\rho ~dv^2 +2dx^2+2e^{-2x}~dy~du,$ & \\
 &  $ L^{2}_{\pm} = e^{-2x} \partial_{_\pm} y,$  & $B=e^{-2x}~du\wedge dy$ & $I=\beta_{_3} \partial_{_u}$\\
 &  $ L^{3}_{\pm} = -u^2 e^{-2x} \partial_{_\pm} y-2u \partial_{_\pm} x + \partial_{_\pm} u$  &  & \\
  &  $L^{4}_{\pm} = \partial_{_\pm} v,$  &  & \\
\\
 {$A_{5,3}$} & $ L^{1}_{\pm} =  \partial_{_\pm} x +\frac{z^2}{2} \partial_{_\pm} u,$ & $ds^{2}=-2\lambda_1\big(dx~du+dx~dv-dy~dz$ & \\
 & $L^{2}_{\pm} =\partial_{_\pm} y + v z \partial_{_\pm} u -\frac{v^2}{2} \partial_{_\pm}z,$  & ~$+\frac{z^2}{2}~du dv\big) +\lambda_2(du+dv)^2+\rho dz^2,$ & \\
 & $L^{3}_{\pm} = z \partial_{_\pm} u -v \partial_{_\pm} z,$ &  $B=\frac{1}{2} z^2~du\wedge dv$ & $I=\beta_{_1} \partial_{_x}+\beta_{_2} \partial_{_y}$\\
  &  $L^{4}_{\pm} = \partial_{_\pm} u + \partial_{_\pm} v,$  &  & \\
   & $L^{5}_{\pm} = \partial_{_\pm} z$  &  & \\\hline \hline
\end{tabular}}}
\end{center}
\vspace{-3mm}
{\small $^{\ast}~\bar{\phi} = (\varphi_{_1} + \varphi_{_2})/2$.}
\\
Indeed, the vector field \eqref{3.7} satisfies equation \eqref{2.39}.
Using \eqref{3.7}, equations \eqref{2.37}, \eqref{2.38}, \eqref{2.40} and \eqref{2.41} with
the metric \eqref{3.5} and $B$-field \eqref{3.6} are satisfied if the constants $\beta_{_1}$, $\beta_{_2}$ and $\beta_{_3}$ are considered to be zero.
Thus, we showed that the background of the $H_4$ WZW model with the vector field $I=\beta_4~\partial_{_v}$ is a solution for the GSEs.

For the sake of clarity, the results obtained in this section are summarized in Table 1;
we display the backgrounds corresponding to the WZW models on real Lie groups up to dimension five,
together with the vector field and corresponding left-invariant one-forms.

\section{Conclusions}
\label{Sec.IV}
In the present work we have investigated the GSEs for the WZW model in such a way that
to simplify the GSEs we have imposed the conditions for vanishing of the one-loop beta function equations.
Then, by introducing an {\it Ansatz} for the vector field $I$, we have proved that the Killing equation ${\cal L}_{_{I}} G_{_{\mu \nu}}=0$ is held.
Furthermore, we introduced a generalized Killing vector field in terms of $I$ and observed that a necessary condition for
having a solution to the GSEs is that this generalized vector be light-like.
We have classified some WZW models based on real Lie groups up to dimension five as solutions of the GSEs.
As shown, the $SO(3)$ and $SO(3) \otimes SO(2)$ Lie groups did not have light-like vectors, that is, did not satisfy equation \eqref{2.31}.
Accordingly, the WZW models built on these groups did not admit the GSEs.
However, we think that our results in the present work can still provide insights into the GSEs.
The construction introduced in the present article for solving the GSEs admits all backgrounds built on
the Lie groups that satisfy the usual supergravity equations.
If the corresponding dilaton field is not constant, equation \eqref{2.21} must also be added to equations \eqref{2.23}-\eqref{2.28}.
Furthermore, if the backgrounds are not constructed on Lie groups, the vector field $I$ no longer follows from equation \eqref{2.42}, and
instead must be constructed from a linear combination of the Killing vectors of the metric (see \cite{TTMP.eghr}).

On the other hand, examining the solutions of GSEs under the T-duality is itself a very interesting problem, as this issue has been investigated in some articles \cite{{M.Y.E.Colgain},{hlavaty1},{hlavaty2},{hlavaty3},{TTMP.eghr}}.
Regarding the interpretation of the solutions of the GSEs to the formal T-duality, we have lately shown that non-Abelian T-dual target spaces of the Yang-Baxter deformed backgrounds of the $H_4$ WZW model \cite{Epr1} admit the GSEs. In fact, the solutions of the GSEs are, under the T-duality, preserved.
The results of this work are still under consideration.
Finally, it would be interesting to write down the GSEs on supermanifolds, and followed by find new solutions for the resulting equations
based on the WZW models on Lie supergroups in low dimensions \cite{{ER.super1},{EPR}}.
For this purpose, one must pay attention to the fact that the vector field $I$ should be considered with two degrees, even and odd.
We intend to address this problem in the future.


\subsection*{Declaration of competing interest}

The authors declare that they have no known competing financial
interests or personal relationships that could have appeared to influence
the work reported in this paper.

\subsection*{Acknowledgements}

The authors are greatly indebted to the anonymous referee for the constructive
comments to improve the presentation of this work.
This work has been supported by the research vice
chancellor of Azarbaijan Shahid Madani University under research fund No. 1402/537.


\subsection*{Data availability}

No data was used for the research described in the article.


\end{document}